\documentclass[twocolumn,epjc3]{svjour3}
\usepackage[T5,T1]{fontenc}
\usepackage{graphicx} 
\journalname{Eur. Phys. J. Plus}

\usepackage{lineno}
\usepackage[numbers]{natbib}

\begin{document}

\title{Citizen Science for IceCube: \textit{Name that Neutrino}}

\onecolumn
\author{R. Abbasi\thanksref{loyola}
\and M. Ackermann\thanksref{zeuthen}
\and J. Adams\thanksref{christchurch}
\and S. K. Agarwalla\thanksref{madisonpac,bhubaneswar}
\and J. A. Aguilar\thanksref{brusselslibre}
\and M. Ahlers\thanksref{copenhagen}
\and J.M. Alameddine\thanksref{dortmund}
\and N. M. Amin\thanksref{bartol}
\and K. Andeen\thanksref{marquette}
\and G. Anton\thanksref{erlangen}
% ...
}

\authorrunning{IceCube Collaboration}


\institute{
% ...
Universit{\'e} Libre de Bruxelles, Science Faculty CP230, B-1050 Brussels, Belgium \label{brusselslibre}
% ...
\and Department of Physics, Loyola University Chicago, Chicago, IL 60660, USA \label{loyola}
\and Dept. of Physics and Astronomy, University of Canterbury, Private Bag 4800, Christchurch, New Zealand \label{christchurch}
% ...
\and Niels Bohr Institute, University of Copenhagen, DK-2100 Copenhagen, Denmark \label{copenhagen}
\and Dept. of Physics, TU Dortmund University, D-44221 Dortmund, Germany \label{dortmund}
% ...
\and Erlangen Centre for Astroparticle Physics, Friedrich-Alexander-Universit{\"a}t Erlangen-N{\"u}rnberg, D-91058 Erlangen, Germany \label{erlangen}
% ...
\and Dept. of Physics and Wisconsin IceCube Particle Astrophysics Center, University of Wisconsin{\textemdash}Madison, Madison, WI 53706, USA \label{madisonpac}
% ...
\and Department of Physics, Marquette University, Milwaukee, WI 53201, USA \label{marquette}
% ...
\and Bartol Research Institute and Dept. of Physics and Astronomy, University of Delaware, Newark, DE 19716, USA \label{bartol}
% ...
\and Deutsches Elektronen-Synchrotron DESY, Platanenallee 6, D-15738 Zeuthen, Germany \label{zeuthen}
\and also at Institute of Physics, Sachivalaya Marg, Sainik School Post, Bhubaneswar 751005, India\label{bhubaneswar}
% ...
}

\date{Received: date / Accepted: date}

\maketitle

\begin{abstract}
\textit{Name that Neutrino} is a citizen science project where volunteers aid in classification of events for the IceCube Neutrino Observatory, an immense particle detector at the geographic South Pole. From March 2023 to September 2023, volunteers did classifications of videos produced from simulated data of both neutrino signal and background interactions. \textit{Name that Neutrino} obtained more than 128,000 classifications by over 1,800 registered volunteers that were compared to results obtained by a deep neural network machine-learning algorithm. Possible improvements for both \textit{Name that Neutrino} and the deep neural network are discussed.

\keywords{neutrino \and citizen science \and Zooniverse}
\end{abstract}

\section{Introduction}

\textit{Name that Neutrino}~\cite{ntnwebsite} is a citizen science project that seeks input from the public to aid in classification of neutrino events for the IceCube Neutrino Observatory. Citizen science provides a powerful tool for advancing both science and education and outreach. Motivated volunteers learn more about cutting-edge research and then perform analyses based on visual scans of data. The ultimate goal is to design, develop and implement an online experience that allows novices to contribute to ongoing research. Results are aggregated to reach statistics not possible within the traditional research community, enabling new insights and discoveries. This paper gives a summary of the IceCube's promising first attempt to engage people outside the collaboration to analyze data using the citizen science approach.

\textit{Name that Neutrino} is hosted on Zooniverse~\cite{zooniverse}, the largest web-based research platform of its kind with over a million volunteers world-wide. Zooniverse has established the power of the citizen science approach. For example, Galaxy Zoo, a Zooniverse project that works on galaxy shape classification, has had more than 10,000 volunteers, resulting in over 60 publications \cite{geron_galaxy_2023}. Their work, much like IceCube's discussed here, is related to pattern recognition and capitalizes on the keen ability of humans to see things that currently remain difficult to identify with computers, even with advances in machine-learning algorithms. However, \textit{Name that Neutrino} is one of the few projects on Zooniverse to include videos rather than static images.

Identifying a research question and developing and implementing the tools needed for \textit{Name that Neutrino} has been a long process. The first attempt at an IceCube citizen science project began in June 2016 at a six-week internship program for high school students \cite{quarknet}. This group produced much of the background material to introduce the IceCube project to the citizen users but the effort was limited to displaying data in static images at that time.  As described in more detail the next section, the IceCube Neutrino Observatory is a cubic kilometer array of 5,160 light sensors (Digital Optical Modules or DOMS) embedded in the South Pole ice.  From  the amount and time sequence of recorded light, the energy and direction of the incident particle can be reconstructed. A video of the time sequence of the data for each event is much more informative, especially for novices. 

The ability to include videos was eventually implemented by Zooniverse, and we conceived of the idea to look at results from people classifying IceCube events and see how they compared to current state-of-the-art machine-learning algorithms. A formal Zooniverse launch was completed in 2023 which included a rigorous approval process in order to be featured to the Zooniverse community. Videos for \textit{Name that Neutrino} were produced from Monte Carlo simulations of trigger-level IceCube data which included significant noise and many ambiguous events. Trigger-level refers to the as-collected data before any quality cuts or noise reduction techniques have been applied. The primary motivation for using trigger-level data was to inspect and compare the performance of both citizen users and a deep neural network (DNN) machine-learning algorithm at the most challenging level.

\begin{figure}
    \centering
    \includegraphics[angle=0,scale=.1]
    {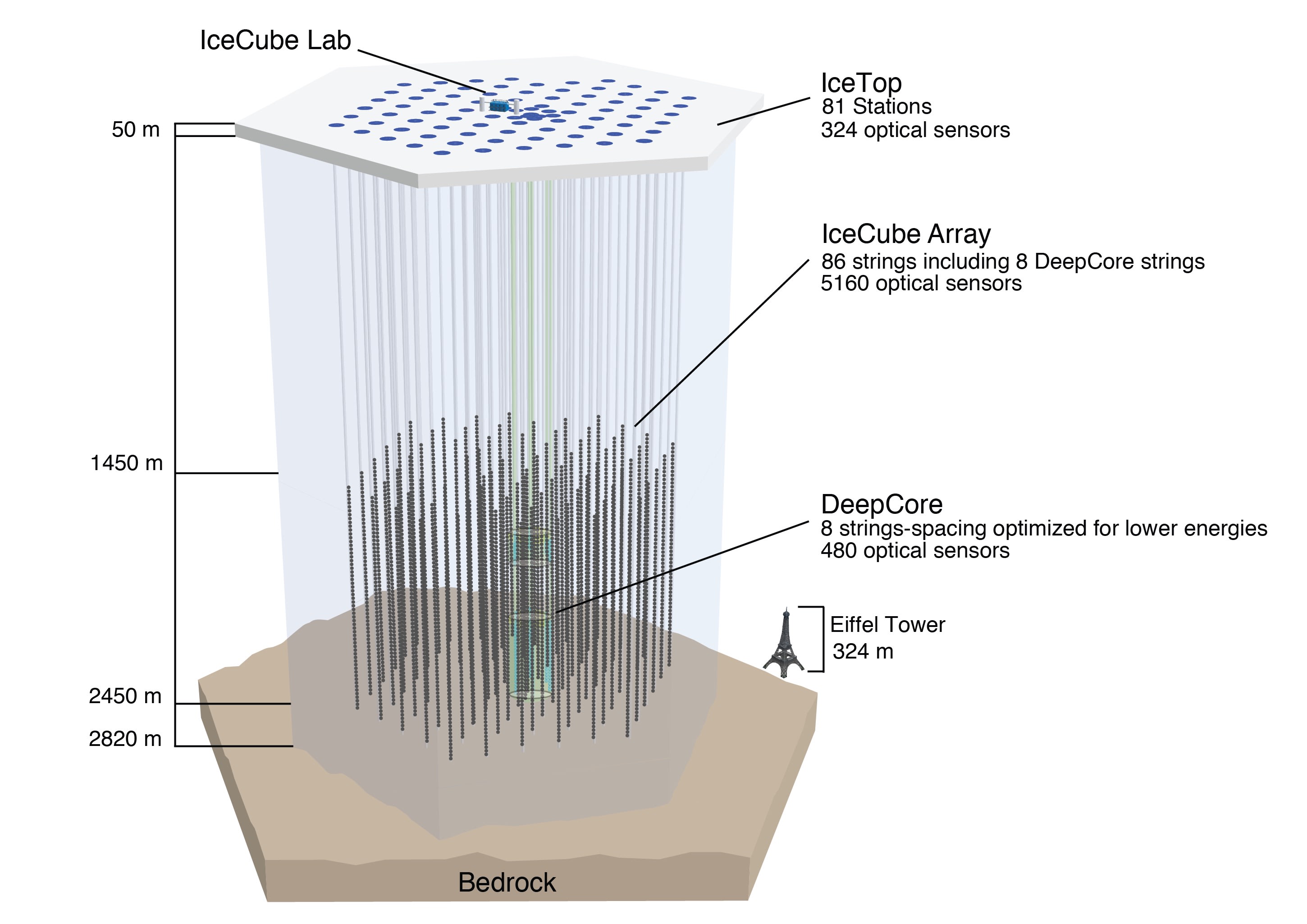}
    \caption{The IceCube Neutrino Observatory.}
    \label{fig:icecube}
\end{figure}

\section{IceCube Neutrino Observatory}

The IceCube Neutrino Observatory~\cite{icecube}, located at the geographic South Pole,  has a multi-faceted and growing science scope. Research topics include astrophysics, particle and fundamental physics, glaciology, and  more, with exciting results identifying the first high-energy neutrino sources~\cite{science2, science3, science4}, leading limits on certain classes of dark matter candidates~\cite{darkmatter} and measurements of neutrino oscillations at energies beyond those reachable in dedicated reactor and accelerator experiments~\cite{osc}. IceCube consists of a hexagonal array of DOMs that instrument a cubic kilometer of ice at depths between 1.5 to 2.5 kilometers below the surface as shown in Fig.~\ref{fig:icecube}.

When a neutrino interacts with with the rock below or the ice near or within the IceCube array, the resulting high-energy secondary particles emit Cherenkov light, some of which is detected by the DOMs. The light pattern (or topology) will depend on the neutrino characteristics (flavor, direction, and energy) and the type of interaction~\cite{allsky}. In the work presented here, IceCube cannot distinguish between neutrinos and anti-neutrinos. Since there are three neutrino types, and two channels of neutrino interaction relevant to this work, there are in principle six different options. In practice most of the recorded events fall into two broad categories depending on the neutrino flavor and how it interacted--- either a charged or neutral current interaction. The two main types of topologies seen in IceCube are tracks and cascades. Tracks are produced by muons originating from muon neutrino charged current interactions or cosmic-ray induced air showers. Muons with sufficient energy can propagate large distances in the ice resulting in a linear light pattern. Cascades are produced by particle showers induced by all  neutral current neutrino interactions, as well as electron- and tau  charged current neutrino interactions. These particle showers evolve over distances of approximately 10 meters in the ice, resulting in a roughly spherical outward going light pattern.

Unfortunately, at least for those only interested in neutrinos, there is an overwhelming background--- a steady rain of muons produced by cosmic-ray interactions above the detector in the Earth's atmosphere from the southern hemisphere. Cosmic-ray interactions in the atmosphere also produce neutrinos, which are identified at the rate of about one per million background events. Roughly one in a few hundred of the neutrinos which interact in the detector is from an astrophysical source rather than a cosmic-ray interaction in the Earth's atmosphere. IceCube sees about 3000 events a second, almost entirely background events from cosmic-ray induced muon tracks~\cite{Aartsen_2017}.

\begin{figure}
    \centering
    \includegraphics[angle=0,scale=0.65]{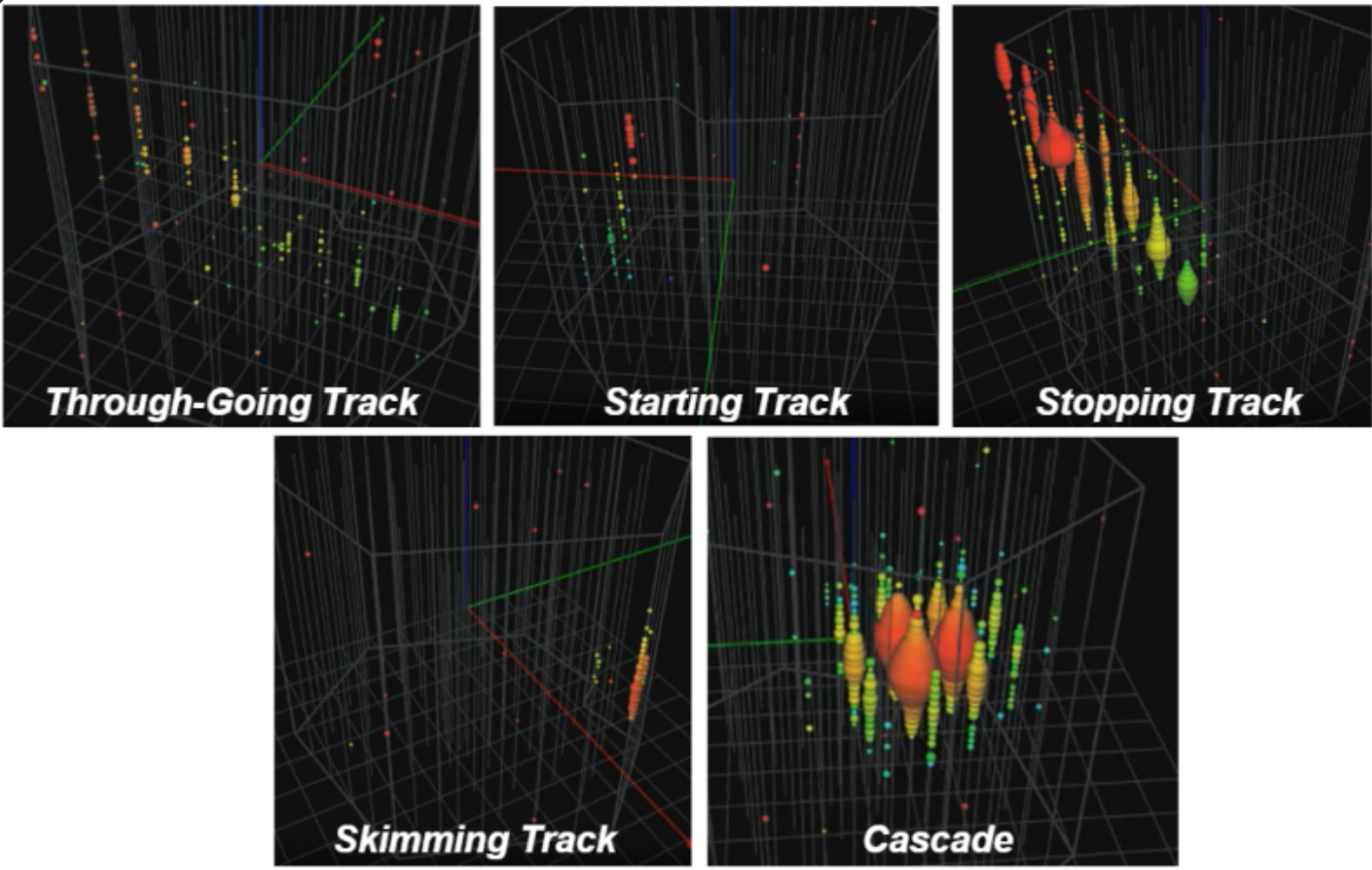}
    \caption{Examples of the five signal topologies used in this work. The color indicates the arrival time at individual DOMs with red happening first, then yellow, green, and blue last. The size of each bubble is related to the light detected by the DOM.}
    \label{fig:top}
\end{figure}

Many investigations of IceCube data begin by separating events into the two broad categories, tracks or cascades.  The track events can be further subdivided into groups as shown in Fig. \ref{fig:top}. Each colored ``bubble” in these event displays represents a DOM that detected light. The size of each bubble is related to the total Cherenkov light detected. The colors indicate the relative time the light was received, with red first and blue last. Through-going tracks start and end outside of the detector and therefore the bubbles transverse the volume. Starting tracks begin inside the detector and move outward, identified by redder bubbles on the interior and bluer bubbles toward the edge. Starting tracks and cascades are some of the most interesting topologies for IceCube since they are only produced by neutrino signal events. However, classification is further complicated since starting tracks are usually accompanied by a cascade from the hadronic part of the charged current interaction and starting tracks turn into cascades as the track length approaches zero. Stopping tracks start outside of the detector and stop inside, with redder bubbles near the edge and bluer bubbles on the interior. Skimming tracks are events all energy loss outside of the detector volume; the detected light will be near the outer volume of the instrumented ice only making it difficult to reconstruct the energy and direction of the incident particle.

IceCube has developed a DNN to classify events that returns a probability of an event being one of the five event topologies: a cascade, or a through-going, starting, stopping, or skimming track. In general, DNNs are a class of machine-learning algorithms that attempt to mimic the brain by learning from known examples. The specific architecture of the DNN used in this work is described in Ref. \cite{Kronmueller:2019jzh} and it was inspired by Google’s InceptionResnet architecture \cite{szegedy2016inceptionv4}. Each event is interpreted as a 4D image with three spatial detector dimensions and one feature dimension that contains information about the time and charge recorded at each DOM. 

Training samples for DNNs usually have a majority of examples that are clear enough for the algorithm to learn. For this particular DNN, the  training used Monte Carlo simulations of events after cuts to the trigger-level data that suppressed noise, background, and ambiguous events. The Monte Carlo truth value is used to provide the DNN with the correct characterization for the training samples; however, the Monte Carlo truth value for trigger-level events in IceCube is not clear cut. One reason for this is trigger-level events can contain coincident events where there are two or more separate signals present, and therefore the Monte Carlo truth is not single-valued. Furthermore, the aforementioned event topologies have been found by IceCube to be useful classes in later stages of our data processing  pipeline. They are based on knowledge about the physical processes but they also have some ambiguity (for example how close an event has to be to the detector boundary to count as skimming). On trigger-level, these categories may not be optimal and it was thus a goal of this project to compare the intuition of citizen users against the DNN.

\section{\textit{Name that Neutrino}}
\begin{figure}
    \centering
    \includegraphics[angle=0,scale=0.3]{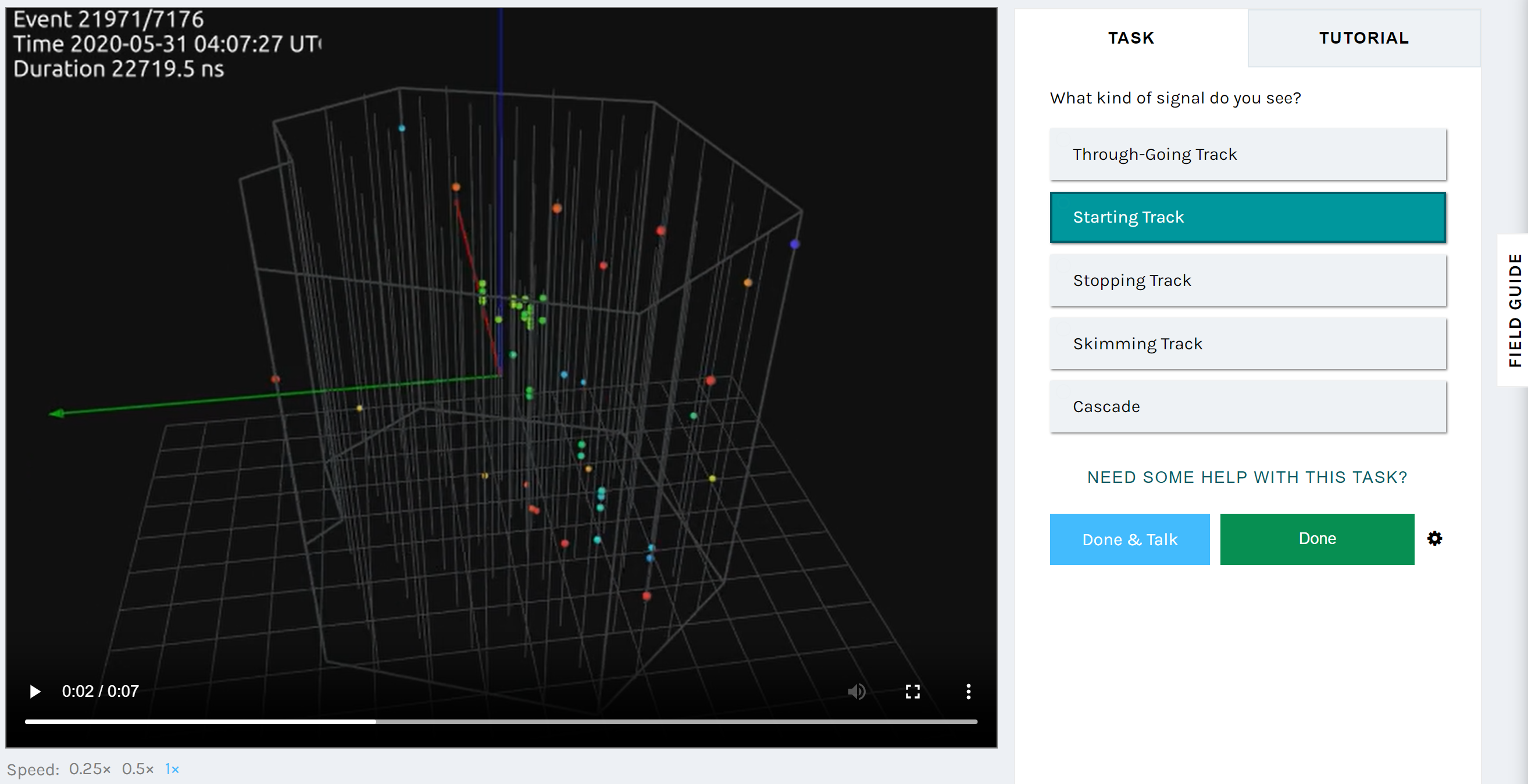}
    \caption{The ``classify” section of the \textit{Name that Neutrino} Zooniverse project. Users are prompted to complete a brief tutorial on their first visit and presented with a field guide with FAQs.}
    \label{fig:class}
\end{figure}

\textit{Name that Neutrino} is available in English, Spanish, and German to anyone in the world with internet access. After selecting ``classify,” first time visitors are prompted with a brief tutorial on how to perform the task, and a field guide with frequently asked questions. Once the tutorial is completed, users have access to a random event from the sample of 4,273 videos. The ``classify” section shown to users can be seen in Fig. \ref{fig:class}. They are able to replay the 7-second video as many times as desired and adjust the playback speed. Then the users must choose from one of the topologies described previously and shown in Fig. \ref{fig:top}.

The 4,273 simulated trigger-level events chosen for \textit{Name that Neutrino} were randomly selected to produce a uniform distribution in the log of the energy---the number of DOMs detecting light scales with energy. With trigger-level events, it is expected to have events that are difficult to classify, especially events with energies close to the detector threshold. An artificial enhancement of electron neutrino events was also implemented to ensure there was a variety of topologies. The enhancement was needed due to the lower trigger rate of electron neutrinos compared to much longer lived muons. Cascades produced by electron neutrino interactions cannot travel far and therefore the electron neutrino must interact inside of the detector. Muon neutrinos on the other hand can interact far from the detector producing long-lived muons that make it inside the detector.

Videos (10 frames per second) were produced with the IceCube event display software Steamshovel~\cite{steamshovel} showing 3D  visualizations of each event with the detector rotating by 5.3 degrees per second. The choices for the videos were chosen to provide a variety of viewpoints with adequate video quality after compression to fit within the Zooniverse file limit of 1 MB per event file. The videos were uploaded as a subject-set to work with a uniquely designed Zooniverse classification workflow \cite{WarrickElizabethHildaStern2023BaCS}.

The Zooniverse approval process requires reviews and beta tests; \textit{Name that Neutrino} completed one internal Zooniverse review and two beta tests that provided feedback from citizen users who suggested improvements and assessed the feasibility of the project. Zooniverse approved and officially launched \textit{Name that Neutrino} in March 2023; after 3 months each of the videos were classified 15 times (by 15 different users) resulting in 64,095 classifications. The video repetition is standard practice for Zooniverse projects and is called the retirement limit. When an individual user is classifying events for \textit{Name that Neutrino}, they will be shown a randomly chosen video that has not yet met the retirement limit and was not previously seen by that user. In June 2023, the retirement limit was increased to 20 and all of the videos were then classified 5 more times. By September 2023, the new retirement limit of 20 was reached for all videos, resulting in 85,460 classifications. As of December 2023, there are over 128,000 classifications and over 1,800 registered volunteers for \textit{Name that Neutrino} who continue to work toward the higher retirement limits.

\section{Results}
After the classification of all 4,273 events, a maximum score is calculated for each event. The maximum score for either the Zooniverse citizen users or for the DNN is the probability value of the most-likely category. For the users, the maximum score is calculated as the fraction of the total votes that the category with the most votes received. This is a proxy for confidence as it shows the level of agreement between users. A maximum score of 0.2 is the minimum possible value and indicates no preference for any classification. A maximum score of 1 means that every user chose the same classification. For the DNN, the maximum score is defined to be the highest probability among the five categories provided by the DNN.

\begin{figure}
   \centering
  \includegraphics[angle=0,scale=0.222]{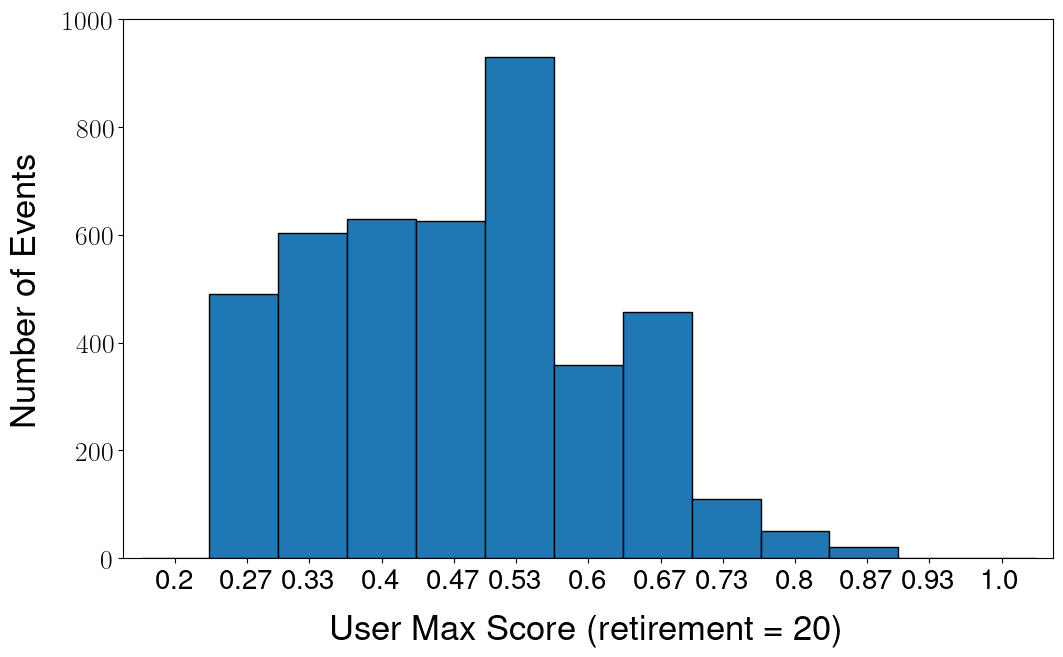}
  \includegraphics[angle=0,scale=0.6]
  {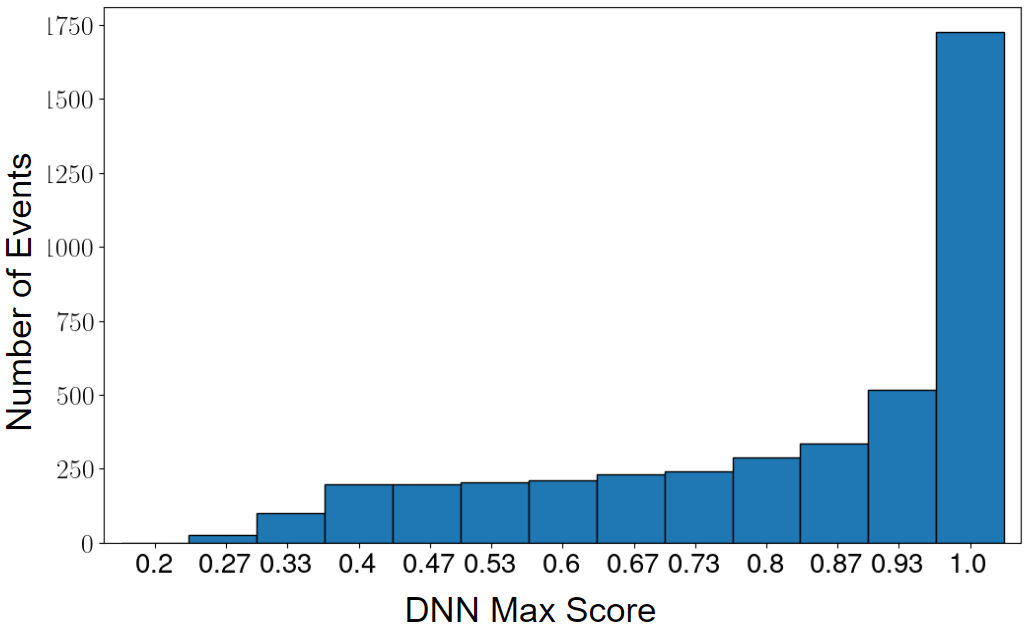}
   \caption{The distribution of maximum scores for the citizen science users (left) and the DNN machine-learning algorithm (right).}
    \label{fig:maxscore}
\end{figure}

The distribution of maximum scores are shown in Fig. \ref{fig:maxscore} for a retirement limit of 20 classifications per event for users (left), and for the DNN (right). For users, the distribution is broadly peaked around 0.53 with the majority of events achieving a maximum score less than 0.5. The low user scores signify disagreements between users and are likely related to the difficulty of classifying trigger-level events. There are no events with a maximum score below 0.25, indicating that there is always at least a weak preference for a classification. The distribution of DNN maximum scores peaks sharply at 1, indicating a high confidence in the classification. It is important to point out that high confidence does not necessarily correlate with accuracy. The DNN was applied to data with more information than the training set, and could preferentially select the same classification for events outside its original scope.

\begin{figure}
    \centering
    \includegraphics[angle=0,scale=0.28]{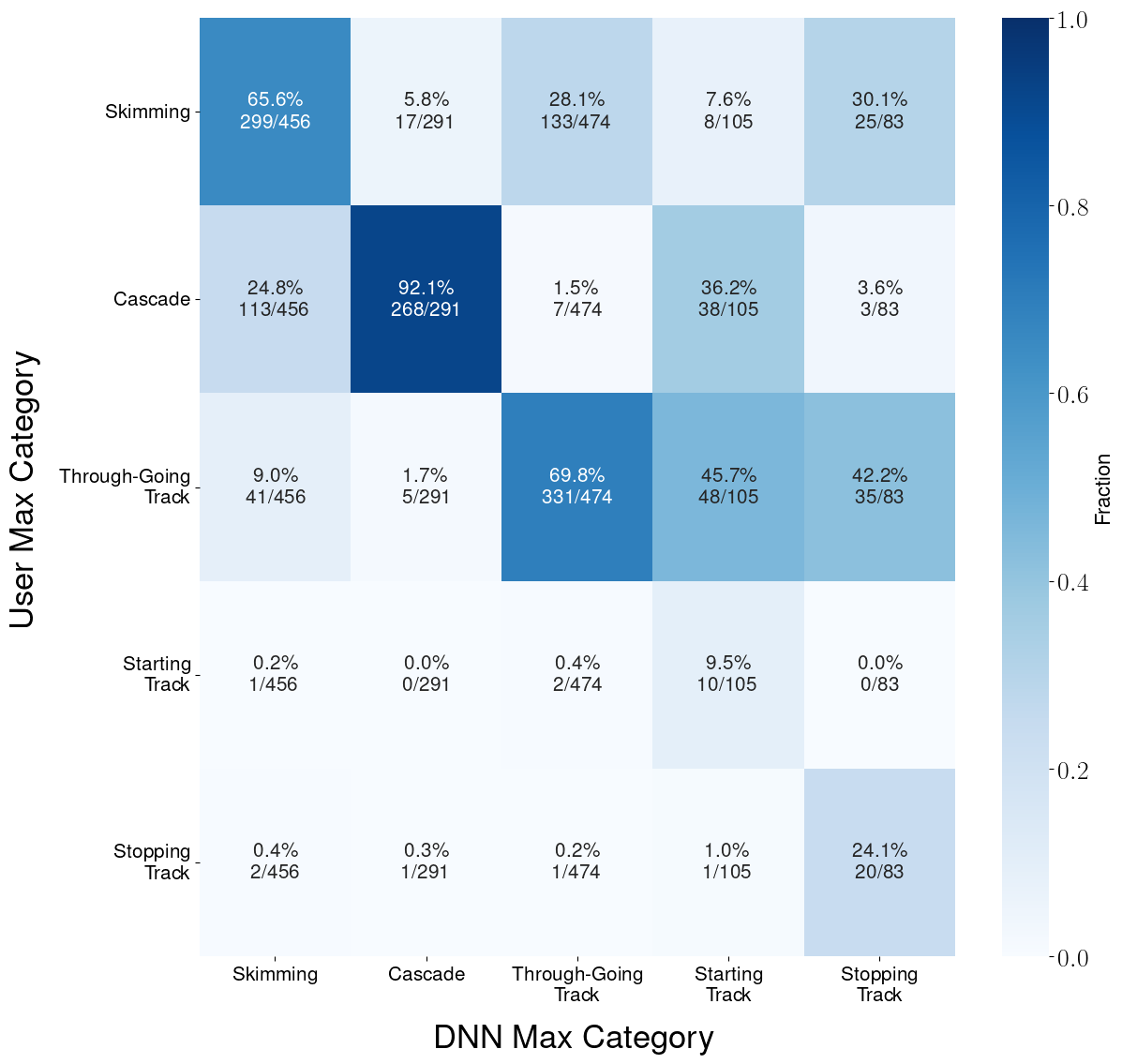}
    \caption{A confusion matrix showing the agreement along the diagonal between the user and DNN maximum categories, normalized to the maximum DNN category. Only user events with a maximum score of 0.55 and above are included.}
    \label{fig:retire20}
\end{figure}

The confusion matrix shown in Fig. \ref{fig:retire20} compares the maximum categories chosen by the DNN to the maximum categories chosen by the citizen science users. To better compare the broad, low-confidence user distribution with the sharply peaked, high-confidence DNN results, we applied a cut to remove user data with a maximum score of below 0.55. The confusion matrix counts the number of events in each category chosen by users compared to and normalized to the DNN maximum category. The columns display events in the associated DNN maximum categories and the rows include events in the corresponding user maximum categories. The diagonal values represent agreement between the DNN and the users, with the largest agreement occurring with cascades at 92.1\%. The disagreement between the users and the DNN is represented by off-diagonal values. For example, in the top right corner of Fig. \ref{fig:retire20}, there were 25 events (30.1\%) where the users chose skimming out of the 83 total stopping tracks as classified by the DNN. Similarly, users chose another 35 of those 83 events (42.2\%) to be through-going tracks instead of stopping tracks. Some of the confusion shown in Fig. \ref{fig:retire20} comes from differences in the training methods for the users and the DNN. The DNN was trained on around 13 million events with concrete definitions for each classification and the users were given a qualitative explanation of each classification and one example image per classification. Finally, though this confusion matrix is normalized to the DNN maximum category, this does not imply that is the correct characterization of the event. The matrix simply shows levels of agreement between users and the DNN results.

\begin{figure}
   \centering
  \includegraphics[angle=0,scale=0.22]{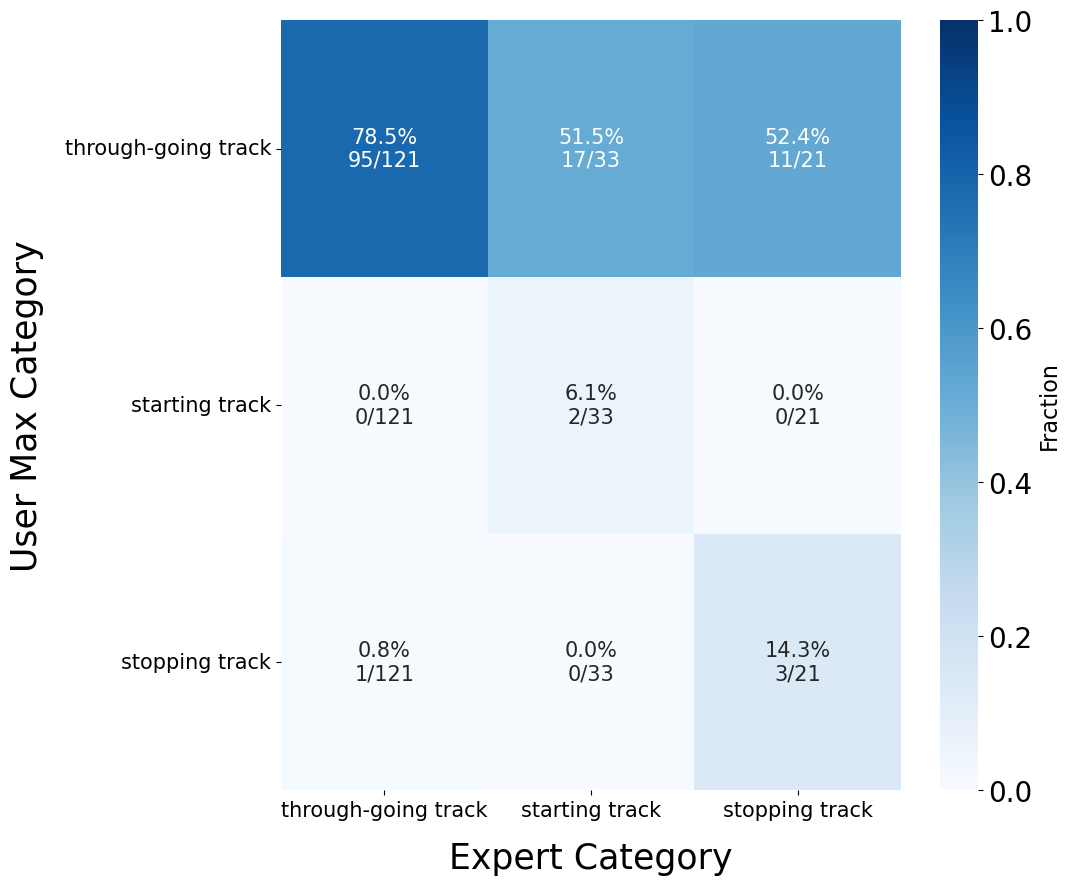}
  \includegraphics[angle=0,scale=0.22]{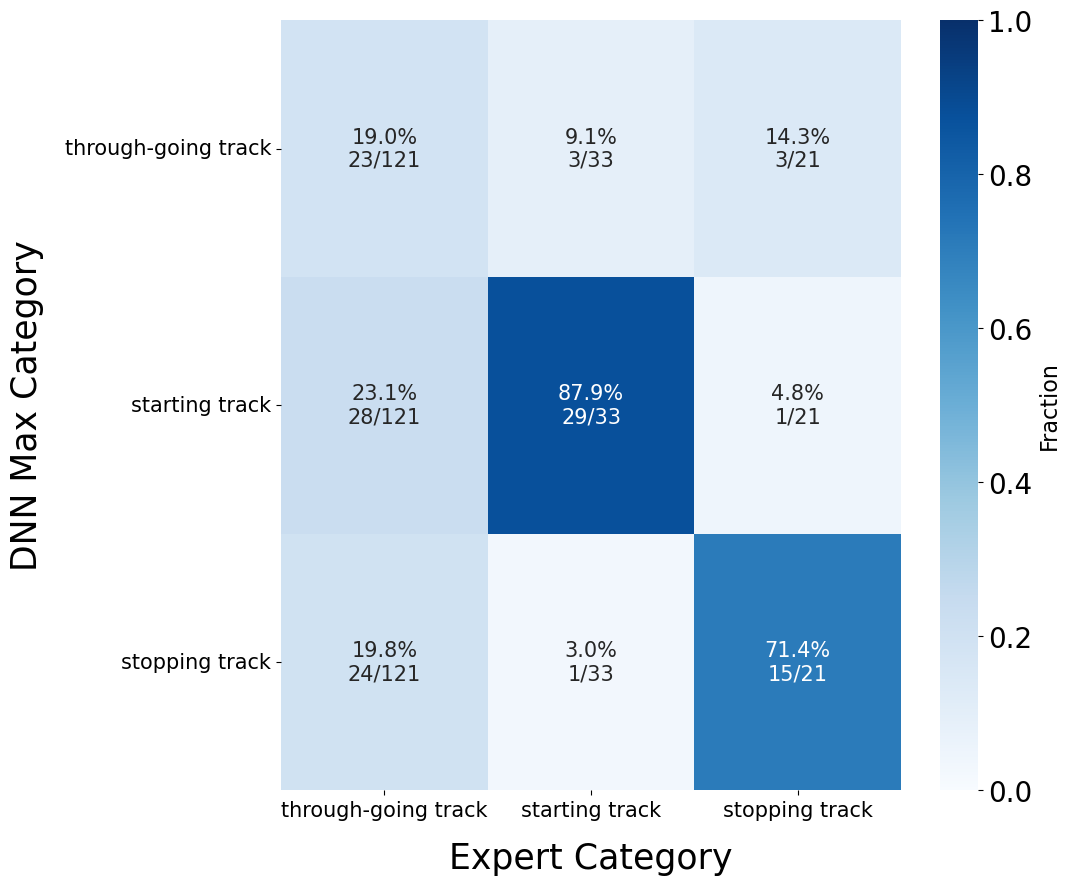}
   \caption{Confusion matrices showing both user (left) and DNN (right) results normalized to the ``expert-by-eye” value for through-going tracks, starting tracks, and stopping tracks. Only the off diagonal values from Fig. \ref{fig:retire20} were used in order to explore disagreeing scenarios. Note it is possible for these events to be classified as cascades or skimming tracks.}
    \label{fig:expert}
\end{figure}

Since the events are trigger level, ``expert-by-eye” values were produced by members of the \textit{Name that Neutrino} team as a substitute for the Monte Carlo truth. Specifically, we explored the off-diagonal starting, stopping, and through-going track events where users and the DNN disagreed on the classification. The comparison of the user and DNN classifications with the ``expert-by-eye” classifications are shown in the left and right sides of Fig. \ref{fig:expert} respectively. Again, diagonal values represent agreement and off-diagonal values represent disagreement, but now in comparison to the ``expert-by-eye”. Note the columns do not sum to 100\% since it is still possible for selected events to be classified as cascades and skimming tracks. Future work will explore these scenarios.
 
Though the number of events has been significantly reduced, there are some large, noteworthy differences between the performance of the users and the DNN. For through-going tracks, users agreed with the expert category 78.5\% of the time compared to 19.0\% for the DNN. This could be an indication that the users are better at identifying through-going tracks or that users were more inclined to pick through-going track because they took that option much more often than other categories. Alternatively, the DNN agreed with the expert category more often for the starting and stopping tracks than the users. This may be due to difficulty in identifying the edges of the detector in the views available to the users. More work is needed to understand the results shown in Fig. \ref{fig:expert}.

\section{Conclusion and Future Work}
After seven months of collecting data, the \textit{Name that Neutrino} project has provided insights into data classification for both citizen users and an IceCube specific DNN machine-learning algorithm. There was more agreement between the users and the DNN for events classified as cascades and less consistency for events classified as starting, stopping, or through-going tracks. The comparison of user and ``expert-by-eye” classifications indicates that additional training is needed for the users. More work should be done in order to fully understand the DNN results compared to the ``expert-by-eye” classifications. This initial study has demonstrated the feasibility of using the citizen science approach to classify IceCube data but that more work is needed to establish the validity of the user results.

Future improvements to \textit{Name that Neutrino} should include increasing the rotation and sharpening the edges of the detector to help improve the identification of starting and stopping tracks by the users. Since the task of classifying events at the trigger-level was rather challenging, future iterations of \textit{Name that Neutrino} could include cleaner events where it would be possible to compare the Monte Carlo truth values to the user results. Or alternatively, more work could be done to optimize the Monte Carlo truth values at trigger-level in order to compare to the results presented here. Future citizen science projects could involve the use of real data rather than simulation data to separate coincident events, identify new classes, or search for possible biases in DNN performance that come from the differences between simulation and data.

\textit{Name that Neutrino} demonstrated that citizen science is a powerful tool for public engagement for IceCube. It was not clear that there would be interest in looking at the rather abstract data, especially compared to more readily identifiable astronomical optical telescope images. Engaging with over 1,800 members of the general public with the IceCube project through 128,000 classifications and over 600 discussion board posts certainly counts as a successful start.

\section{Acknowledgements}

The IceCube collaboration acknowledges significant contributions to this manuscript from Christina Love and Jim Madsen.

The authors gratefully acknowledge the support from the following agencies and institutions: USA {\textendash} U.S. National Science Foundation-Office of Polar Programs,
U.S. National Science Foundation-Physics Division,
U.S. National Science Foundation-EPSCoR,
U.S. National Science Foundation-Office of Advanced Cyberinfrastructure,
Wisconsin Alumni Research Foundation,
Center for High Throughput Computing (CHTC) at the University of Wisconsin{\textendash}Madison,
Open Science Grid (OSG),
Partnership to Advance Throughput Computing (PATh),
Advanced Cyberinfrastructure Coordination Ecosystem: Services {\&} Support (ACCESS),
Frontera computing project at the Texas Advanced Computing Center,
U.S. Department of Energy-National Energy Research Scientific Computing Center,
Particle astrophysics research computing center at the University of Maryland,
Institute for Cyber-Enabled Research at Michigan State University,
Astroparticle physics computational facility at Marquette University,
NVIDIA Corporation,
and Google Cloud Platform;
Belgium {\textendash} Funds for Scientific Research (FRS-FNRS and FWO),
FWO Odysseus and Big Science programmes,
and Belgian Federal Science Policy Office (Belspo);
Germany {\textendash} Bundesministerium f{\"u}r Bildung und Forschung (BMBF),
Deutsche Forschungsgemeinschaft (DFG),
Helmholtz Alliance for Astroparticle Physics (HAP),
Initiative and Networking Fund of the Helmholtz Association,
Deutsches Elektronen Synchrotron (DESY),
and High Performance Computing cluster of the RWTH Aachen;
Sweden {\textendash} Swedish Research Council,
Swedish Polar Research Secretariat,
Swedish National Infrastructure for Computing (SNIC),
and Knut and Alice Wallenberg Foundation;
European Union {\textendash} EGI Advanced Computing for research;
Australia {\textendash} Australian Research Council;
Canada {\textendash} Natural Sciences and Engineering Research Council of Canada,
Calcul Qu{\'e}bec, Compute Ontario, Canada Foundation for Innovation, WestGrid, and Digital Research Alliance of Canada;
Denmark {\textendash} Villum Fonden, Carlsberg Foundation, and European Commission;
New Zealand {\textendash} Marsden Fund;
Japan {\textendash} Japan Society for Promotion of Science (JSPS)
and Institute for Global Prominent Research (IGPR) of Chiba University;
Korea {\textendash} National Research Foundation of Korea (NRF);
Switzerland {\textendash} Swiss National Science Foundation (SNSF).

This publication uses data generated via the Zooniverse.org platform, development of which is funded by generous support, including a Global Impact Award from Google, and by a grant from the Alfred P. Sloan Foundation.

\section{Data Availability Statement}
The datasets used and/or analysed during the current study are available at https://zenodo.org/records/10521028 and the videos are available at https://www.zooniverse.org/projects/icecubeobservatory/name-that-neutrino.

\bibliographystyle{sn-mathphys}

\end{document}